\documentclass[aps,final,notitlepage,oneside,twocolumn,nobibnotes,nofootinbib,
superscriptaddress,noshowpacs,centertags]{revtex4-1}

\usepackage[utf8]{inputenc}
\usepackage[english]{babel}
\usepackage{graphicx}
\usepackage{latexsym}
\usepackage{amssymb}
\usepackage{amsmath}
\usepackage{float}
\usepackage{color}

\begin{document}

\title{Cascade Relaxation of the Gravitating Vacuum\\
as a Generator of the Evolving Universe}

\author{V. N. Lukash}\thanks{e-mail: lukash@asc.rssi.ru}
\affiliation{Astro Space Center, P. N. Lebedev Physical Institute, Russian Academy of Sciences, Moscow, 117997 Russia}
\author{E. V. Mikheeva}\thanks{e-mail: helen@asc.rssi.ru}
\affiliation{Astro Space Center, P. N. Lebedev Physical Institute, Russian Academy of Sciences, Moscow, 117997 Russia}

\date{\today}

\begin{abstract}
The cascade relaxation of a polarized vacuum in the expanding Universe is a chain of evolutionary epochs of decreasing its density with the exit of dominant fields (each in its own time) from the initial zero states to nonzero values, from the dominant scalar field in the early Universe to subsequent ones, including the $\Lambda$-term in the modern Universe. The cascade vacuum relaxation creates the entire observable cosmology from the Friedmann model with small perturbations of the metric with non-power-law power spectra from which primordial black holes could have arisen, and gravitational waves over a wide wavenumber range to the formation of dark matter and energy, early galaxies, supermassive black holes, and the large-scale structure of the Universe. An observational model of cascade vacuum relaxation in the early Universe has been build, which contains two constants determined by observational data and does not require information on the potential of the fields. A solution has been obtained for the general relativity vacuum attractor, including, in addition to the two previously mentioned constants, the third constant (not yet limited by observations), which leads to an additional power of density perturbations on a small scale ($k>10\,$Mpc$^{-1}$) in the form of a ``bump'', a two-power-law spectrum, etc.
\end{abstract}

\maketitle 

DOI: 10.1134/S0021364024604718

JETP Letters, 2025, Vol. 121, No. 6, pp.399-410. 

%%%%%%%%%%%%%
\section{INTRODUCTION}
%%%%%%%%%%%%%
Many years have passed since the discovery of the accelerated expansion of the Universe \cite{perlmutter}, but the physical nature of this phenomenon remains unclear, which is manifested in the presence of many hypotheses regarding the nature of the cosmological constant. The general theoretical principle of explaining phenomena by the least number of causes does not work in this case, and the Universe does not have to meet our expectations. The beginning of the dynamic influence of a certain vacuum precisely in the modern era ``pulls'' towards the anthropic principle.

The equation of state of the medium ($E+P \simeq0$), responsible for the accelerated expansion of the Universe, is close to both the vacuum equation $E+P =0$ and the equation of state of the scalar field with the dominant potential term $|E + P| \ll E$. This makes it possible not only to bring the discussion of the nature of the $\Lambda$-term from the classical Einstein cosmological constant to quintessence \cite{quint}, but also to accept the concept of many fields, which is logically associated with the idea of a large number of degrees of freedom (fundamental matter fields) responsible for the polarization of the vacuum in the 
Universe\footnote{We mean that vacuum polarization is the response of the vacuum (the ground or zero state of all fields) to an external force, provided by the non-stationary gravitational potential of matter in the Universe. As a result, the vacuum is deformed and polarized with the appearance of a non-zero density. In the absence of an external force, the vacuum density is zero. Since the external force (gravity) exists, the density of the vacuum is understood as its polarization.} \cite{wing1, wing2, XXX1, book2010, rubakov, kazakov}. Starting from the first insufficiently accurate observational data and replacing them with different field interaction potentials, studies led to different answers for the expected future experimental tests. After 2021 (see \cite{BICEP}), the improved data accuracy allows us in this work to develop an observational model of the early Universe without solving the Friedmann equations and without information on the potential of the fields.

In view of the high accuracy of the observational data, we can already develop an observational model of vacuum relaxation, assuming that the polarization fields were originally in zero states ($\varphi^{(i)}\simeq0$). At some point, one of them (which has the largest mass) turns out to be the dominant field and begins to evolve from zero to some nonzero value, ending its motion in a new state of vacuum (with the density less than the previous one). This process could be repeated with other fields at other times. In fact, the time-evolving gravitational process of {\it cascade vacuum relaxation} (CVR) gives rise to the entire observed cosmology: cosmological density perturbations and gravitational waves, dark matter, dark energy and halos of galaxies, the large-scale structure of the Universe, etc. When peaks (bumps) occur in the power spectra in the Universe, primordial black holes may appear. In the case of a two-power-law spectrum of density perturbations (with an increasing amplitude with $k>10\,$ Mpc$^{-1}$), the conditions in the Universe are formed for the appearance of supermassive black holes \cite{SMBH} and early galaxies \cite{JWST} observed by the James Webb Space Telescope at $z > 10$ before the formation of the large-scale structure of the Universe.

In Section 2, we construct an observational model of vacuum relaxation focused on the density perturbation spectra generated at the first stage of CVR with a dominant scalar field. Section 3 presents a theoretical solution of the general relativity {\it vacuum attractor} (VA), which corresponds to the observational data and has an additional constant. Section 4 is devoted to non-power-law power spectra. The conditions for the solution to enter and exit the VA are discussed in Section 5. The conclusions are presented in Section 6.

%%%%%%%%%%%%%%%%%%%%%%%%%%%%%%%%
\section{OBSERVATIONAL MODEL OF VACUUM RELAXATION}
%%%%%%%%%%%%%%%%%%%%%%%%%%%%%%%%
The Lagrangian density of general relativity consists of two terms: the first includes the metric tensor and its derivatives $\frac{R}{16\pi G}$, where $R=R_\mu^\mu$, $R_{\mu\nu}$ is the Ricci tensor and $G$ is a constant, and the second includes the fundamental degrees of freedom of matter fields and their derivatives: 
%(1)
\begin{equation}
\mathcal L=\mathcal L\left(w^{\left(1\right)},w^{\left(2\right)},...,\varphi^{\left(1\right)},\varphi^{\left(2\right)},...\right),\label{eq1}
\end{equation}
where $w^{\left(1\right)2}\!=\varphi^{\left(1\right)}_{,\mu}\varphi^{\left(1\right)\!,\mu}$, etc.\footnote{Expression (\ref{eq1}) includes all fields of all spins. To simplify the notation, we omit the spatiotemporal indices of material fields and their kinetic members.} 
Further decomposition of the Lagrangian with the separation of individual kinetic members of the fields occurs evolutionarily with a decrease in the energy. The gravitational field described by the metric tensor $g^{\mu\nu}$, is included in function (\ref{eq1}), but its derivatives appear only in the scalar $R$. The separation of this kinetic scalar is the isolation of the gravitational field, which means the appearance of General Relativity.

Energy density includes entropy, particles, and vacuum polarization. The temperature and density of particles decrease during the gravitational expansion of the Universe; as a result, the vacuum polarization begins to dominate. The decrease in the vacuum density occurs through the CVR, i.e., a chain of epochs in each of which one field dominates, while the others remain gravitationally ``frozen'' in a symmetric vacuum state:
%(2) 
\begin{equation}
\mathcal L\rightarrow \mathcal L\left(w,\varphi,...\right)\rightarrow\frac{w^2}{2}-V\left(\varphi,...\right),
\label{perehod}
\end{equation}
where $w^2\!=\!\varphi_{,\mu}\varphi^{,\mu}$ is the kinetic scalar of the field $\varphi\!=\! \varphi^{(1)}\!$ and $V=V(\varphi,...)$ is the potential of all fields. At the first stage, $\varphi$ dominates with the simultaneous separation of its kinetic term from other fields. In the initial (symmetric or ground) state, all fields are equal to zero. The first field that starts from zero is that dominating in the potential (having the maximum mass parameter). The kinetic scalars of the other fields are equal to zero, since they do not evolve due to equations of motion. 

We consider the vacuum relaxation process as the transition of fields in time from a symmetric state $V_0=V(0,0,...)$ to energetically more favorable states: first to the state $V_1=V(\varphi_1,0,...)<V_0$ in the field $\varphi$ (we assume $\varphi>0$ if the other is not indicated), then from the state $V_1$ to the state $V_2=V(\varphi_1,\varphi_2,0,...)<V_1$ in the field $\varphi^{(2)}$, 
etc.\footnote{All states of the potential are positive, because they are larger than the $\Lambda$-term, which is positive from observations: $V_0 > V_1 >V_2 > \ldots > \Lambda > 0$.} 
We assume that the state $V_0$ is created by all vacuum polarization fields, according to which the gravitational expansion process is accelerated in all three spatial directions (often called evolutionary inflation), which makes any spacetime symmetry closer to the Friedmann one.

In the first stage of CVR (exit from the state $V_0$), the field $\varphi=\varphi(x^\mu)$ moves out of the initial state ($\vert\varphi\vert \ll \varphi_1$), being a gravitational source of the evolving spacially flat Friedmann model with small (linear) fluctuations of the metric, which are described by the fields of density perturbations $q=q(x^\mu)$ and gravitational waves $\tilde q_{ij}=\tilde q_{ij}(x^\mu)$ in all coordinates $x^\mu=(t,x^i)$. The solution can be represented as a series in $q$ and $\tilde q_{ij}$:
%(3)
\begin{equation}
\varphi=\varphi\left(N\right)+\alpha\Delta+O\left(\Delta^2\right),
\label{phiN}
\end{equation}
\[
ds^2=\left(1+2\Phi\right)dt^2-a^2\left(1-2\Phi\right)\left(\delta_{ij}+2\tilde q_{ij}\right)dx^idx^j,
\label{metrika}
\]
where $\Delta=q-\Phi$ is the matter velocity potential, $\Phi=\frac{H}{a}\int{\gamma q a\,dt}$ is the gravitational potential, $\tilde q_i^i=\tilde q_{i, j}^j=0$,  $N=N(t)\equiv\ln(a)$, $\alpha =\alpha(N)\equiv\varphi_{,N}$, $H\equiv\dot N$ and $\gamma\equiv-\frac{\dot H}{H^2}$ are functions of the time, $X_{,x}=\frac{\dot X}{\dot x}$ (for functions of the time), the overdot stands for the time derivative; and the spatial indices $i$ and $j$ are raised and lowered by the unit tensor $\delta_{ij}=\text{diag}(1,1,1)$.  

The field $\varphi(N)$ is a source of a homogeneous vacuum background satisfying the Friedmann equations. Solution (\ref{phiN}) arises spontaneously during the gravitational expansion of the spacetime region from the initial homogeneity scale $\sim1/H_i$ followed by an accelerated expansion of the boundaries $\sim a/(a_i H_i)$, which limits the function $\gamma$: $\;\ddot a>0\;\rightarrow\;\gamma<1$.

Of all the fluctuations of the background, there are two gravitational fields $q$ and $\tilde q_{ij}$, which are the scalar ($S$) and tensor ($T$) modes of linear perturbations of the metric, respectively. They are test massless fields of the Friedmann model with actions obtained by the direct expansion of the general action in $q$ and $\tilde q_{ij}$ to the second-order terms \cite{Lukash80a, Lukash80b} and \cite{book2010} (see Appendix):
%(4)
\begin{equation}
\delta^{\left(2\right)}S=\int\left(L+\tilde L\right)\sqrt{-g}d^4x,
\end{equation}
\[
L=\frac{\gamma q_{,\mu}q^{,\mu}}{8\pi G}\!=\frac{\alpha^2q_{,\mu}q^{,\mu}}{2},\;\;\tilde L=\frac{\tilde q_{ij,\mu}\tilde q^{ij,\mu}}{16\pi G}\! =\frac{m_P^2\tilde q_{ij,\mu}\tilde q^{ij,\mu}}{4},
\]
where $m_P=\frac{M_P}{2\sqrt{\pi}}$ with $M_P=\frac{1}{\sqrt G}$. 
Note that the field $q$ enters the velocity potential and has a Newtonian limit, the field $\tilde q_{ij}$ has two polarizations, the speed of the $S$ mode tends to unity, $c_S^{-2}=\frac{w\mathcal L_{,w,w}}{\mathcal L_{,w}}\rightarrow1$, under the condition given by Eq. (\ref{perehod}), and the functions $\alpha$ and $\gamma$ are related as $\gamma=4\pi G\alpha^2$. In addition, it is convenient to introduce the following designations of the derivatives of $H$ and establish useful relations between them:
%(5)
\begin{equation}
\gamma=\!\beta^2\!=\!-\frac{H_{,N}}{H},\;\,\beta=\!\frac{\alpha}{m_P}\!=\!\phi_{,N}\!=\!-\frac{H_{,\phi}}{H},\;\,\varepsilon=\beta_{,\phi},\!
\label{g5}
\end{equation}
where $\phi\equiv\frac{\varphi}{m_P}$ is the dimensionless field corresponding to the dimensional field 
$\varphi$.

The fields have zero mathematical expectations but nonzero variances:
%(6)
\begin{equation}
\langle q^2\rangle=\int_0^\infty{q_k^2\,\frac{dk}{k}},\quad\langle\tilde q_{ij}\tilde q^{ij}\rangle=\int_0^\infty{\tilde q_{k}^2\,\frac{dk}{k}},\label{spectradef}
\end{equation}
where the angle brackets $\langle...\rangle$ denote the averaging of the fields over vacuum states, $q_k$ and $\tilde q_{k}$ are the spectra of cosmological perturbations, and $k$ is the wavenumber. The fields $q$ and $\tilde q_{ij}$ are frozen and the spectra do not depend on the time where the scale of the perturbations exceeds the cosmological horizon ($k < aH$). The slopes (indices) of the spectra and their dependence on the scale and the ratio of the power spectra are given by the expressions
\[
n_k=\frac{d\ln q_k}{d\ln k},\;\, 
\tilde n_k=\frac{d\ln\tilde q_k}{d\ln k},\;\, 
A_k=\frac{d\ln\vert n_k\vert}{d\ln k},\;\, 
r_k=\frac{\tilde q^2_k}{q_k^2},
\]
where $r_k$ is quarter of that in \cite{Plancknew11} by definition.

Cosmological observations \cite{Pl20, BICEP} give the following values in the $k$ wavenumber 
range\footnote{The lower and upper limits of the wavenumber range correspond to the modern event horizon and the scale of dwarf galaxies, respectively. The wavenumber $k$ is measured in inverse megaparsecs for a Hubble constant of $67$ km\, s${}^{-1}$\, Mpc${}^{-1}$.} 
from $2\times10^{-4}$ to $10$\,Mpc${}^{-1}$:
%(7)
\begin{equation}
q_k\!\simeq10^{-5}\!\left(\frac{k_c}{k}\right)^{\!n_c}\!\!\!\!,\;\,
n_c\!=0.0175\pm0.0025,\;\,
r_c\!<\!10^{-2}\!\!,
\label{qk1}
\end{equation}
where $k_c\!=0.05$ is the ``central'' wavenumber.

The cosmological scales corresponded to the event horizons $k = aH$ in the epoch of the generation of the $S$ and $T$ perturbation modes due to causal evolution. The background model of the early Universe was described at that time by the dominant field $\phi$, which determined the function $H=H(\phi)$ and its derivatives $\gamma$, $\varepsilon$, and $\varepsilon_{,N}$. Assuming that the last three functions are much smaller than unity (as confirmed by observations) and retaining only the leading $\gamma$ terms, we obtain the following linear perturbation spectra and indices on scales $k=He^{N}$ (see Appendix and Eq. (\ref{prilq})):
%(8)
\begin{equation}
q_k=\frac{H}{2\pi\vert\alpha\vert},\quad
\tilde q_k=\frac{H}{\pi m_P},\quad
n_k\!=\!-\frac{\varepsilon\!+\!\gamma}{1\!-\!\gamma}\simeq\!-\varepsilon\!-\!\gamma,
\label{qk2}
\end{equation}
where the index $k$ of the model functions is omitted, $(\ln k)_{,N}=1-\gamma\simeq 1$, 
and the ratio of the power spectra is $r_k=4\gamma\simeq-4\tilde n_k$. The comparison of Eqs. (\ref{qk2}) with (\ref{qk1}) shows that the functions $\gamma$, $\varepsilon$, and $\varepsilon_{,N}$ were indeed small, and the $\gamma$ value did not exceed the error of observations: 
$\gamma<0,0025$, $\varepsilon+\gamma=0,0175\pm0,0025$ ($\gamma$ and $\varepsilon$ are taken at $k_c$). Then, the field $\phi$ was near zero and increased, which makes it possible to construct observational model of vacuum relaxation in the form of a series in  $\xi\equiv n_0\phi^2\ll1$:     
%(9)
\[
H\!=\!H_0\!\left(1\!-\!\frac{\xi}{2}\!+\!O\!\left(\xi^2\right)\!\!\right)
\!\!,\;
N\!=\!\ln\!\left(\frac{k}{H}\right)\!=\!\ln\!\left(\frac{k}{k_c}\right)-N_c,\;\;
\]
\[
\gamma=n_0\xi\left(1\!+\!O\!\left(\xi\right)\right)\!,\;\;
\varepsilon=n_0\left(1\!+\!O\!\left(\xi\right)\right)\!,\;\;
\varepsilon_{,N}\!=\!n_0 O\!\left(\gamma\right)\!,
\]
\begin{equation}
q_k=\frac{\mathrm H_0\left(1+O\!\left(\xi\right)\right)}{2\pi n_0\phi}\simeq\frac{\mathrm H_0}{2\pi\beta_c}\left(\frac{k_c}{k}\right)^{n_0}, 
\label{eq14}
\end{equation}
\[
n_k\!=\!-\epsilon\!\left(1\!+\!\xi\left(1\!+\!n_0\right)\!+\!O\!\left(\xi^2\right)\!\right)\!
=\!-n_0\!\left(1\!+\!O\!\left(\xi\right)\!\right)\!, 
\]
\[
A_k\!=\!O\!\left(\gamma\right),\quad
r_k\!=4n_0^2\phi^2\!\left(1+O\!\left(\xi\right)\!\right)\!,\quad
\tilde n_k\!\simeq\!-n_0^2\phi^2\!,
\]
\[
\phi\simeq\!\phi_c e^{n_0\left(N\!+\!N_c\right)}\!\simeq
\!\phi_c\!\left(\frac{k}{k_c}\right)^{\!n_0}\!\!\!,
\]
where $H_0\!$ and $n_0\!$ are constants, $\mathrm H_{(0)}\!\equiv\!\frac{H_{(0)}}{m_P}$, $\beta_c\!=\!n_0\phi_c$, and $N_c\!\simeq\!\ln(\frac{H_0}{k_c})$. The observational model of vacuum relaxation provides a power-law red spectrum in the $k$ wavenumber range from $2\times10^{-4}$ to 10 Mpc${}^{-1}$, does not require information on the potential, and is based on the current observational data:
%(10)
\begin{equation}
\mathrm H_0\!\simeq\!10^{-6}\phi_c,\; N_c\!\simeq\!120\!+\!\ln\phi_c,\; n_0\!\simeq\!0.017,\;\phi_c\!< 3
\label{eq10}
\end{equation}
and $\beta_c\!<0.05$. The function $H$ and its derivatives are determined from current observational data and do not require the solution of the Friedmann equations. They contain two constants $H_0$ and $n_0$ and refer to the time interval when $\varphi<0.8\,M_P$, $\alpha<0.014\,M_P$ and $\xi\simeq\frac{\gamma}{\varepsilon}<0.17$.

The observational model of vacuum relaxation describes the beginning of the first phase of vacuum relaxation, where the field was near zero and increases (remaining less than the Planck mass), which ensures the constancy of the power-law spectral index and the smallness of the $\gamma$-function in the $k$ wavenumber range from $10^{-4}$ to $10$ Mpc${}^{-1}$.\footnote{Other field trajectories (e.g., models with the decreasing field from large to small values) associate the spectral index with the value of $\gamma$ and are inconsistent with the observational data.} 
The constructed observational model of vacuum rlaxation follows from modern data and does not require further verification. For the continuation of the observational model of vacuum relaxation on a small scale ($k > 10$Mpc${}^{-1}$), new data and information on the potential are needed (the counterterms associated with the third constant are within observational errors).

%%%%%%%%%%%%%%%%%%%%%%
\section{VACUUM ATTRACTOR SOLUTION}
%%%%%%%%%%%%%%%%%%%%%%
Since the fields move from zero, we expand the potential in a power series of fields, assuming the first three terms as essential and the rest as small:
%(11)
\[
V=V_0-\frac{m^2\varphi^2}{2}+\frac{\lambda\varphi^4}{4}+...\rightarrow
\]
\begin{equation}
 \rightarrow V_1+m^2\left(\frac{\varphi^2-\varphi_1^2}{2\varphi_1}\right)^2,
\label{3const}
\end{equation}
where $V_{0,1}$ and $\varphi_1=\frac{m}{\sqrt\lambda}$ are positive constants (constant
coefficients of the series are determined by all fields). Since $V$ is an even function of $\varphi$, the field evolves from a symmetric state $V_0=V(0)$ towards positive (or negative) $\varphi$ to an energetically more favorable state $V_1=V(\pm\varphi_1)$ (see Fig.~1):
%(12)
\begin{equation}
V_0=V_1+\frac{m^4}{4\lambda}>V_1>0.
\end{equation}
%  
%%%%%%%%%%%%%%%%%%%%%%%%%%%%%%%%%%%%
\begin{figure}
\centering
\includegraphics[width=0.5\textwidth]{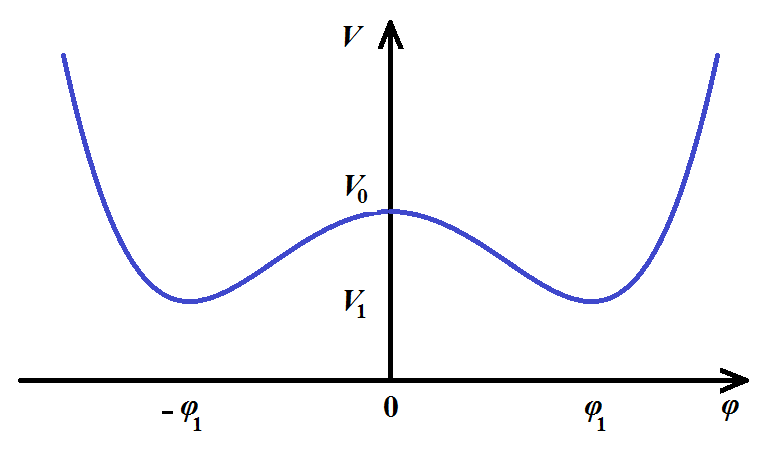}
\caption{Potential $V$ versus the field $\varphi$. The potential is equal to the minimum value $V_1$ at $\varphi=\pm\varphi_1$.}
\label{fig1}
\end{figure}
%%%%%%%%%%%%%%%%%%%%%%%%%%%%%%%%%%%%

When the field rolls down the potential, the height of the vacuum step $V_{01}= V_0-V_1$ depends on two fundamental constants: $V_{01}=\frac{m^4}{4\lambda}=\frac{m^2\varphi_1^2}{4}$. Since Eq.(\ref{3const}) contains three constants, we take the residual vacuum in units of step height, $v_1=\frac{V_1}{V_{01}}$, or the amplitude of the initial vacuum, $\mathrm v_1=\frac{V_1}{V_0}\in(0,1)$ as the third (independent) constant:
%{13}
\begin{equation}
v=v_1\!+y^2,\;\;\mathrm v=\mathrm v_1\!+\frac{y^2}{v_0}=1-\frac{\mathrm n_0\left(1+y\right)\phi^2}{2},
\end{equation}
where $v=\frac{V}{V_{01}}$, $\mathrm v=\frac{v}{v_0}$, $y=1-x^2$, and 
$x=\frac{\varphi}{\varphi_1}=\frac{\phi}{\phi_1}$ are dimensionless variables, $v_0=1+v_1\!=\!(1\!-\!\mathrm v_1)^{-1}$, $\mathrm n_0=\frac{m_P^2m^2}{2\mathrm V_0}$, $\phi_1=\frac{\varphi_1}{m_P}=\sqrt{\frac{2}{v_0\mathrm n_0}}$. At $v_1\!>\!1$, $\mathrm v\!\simeq\!1$ in the range $x\in(0,1)$. 
At $v_1<1$, the function $\mathrm v$ varies from $\mathrm v\simeq 1$ at $y\!\in\!(1,\frac{1}{\sqrt2})$ to $\mathrm v\simeq\mathrm v_1$ at $\!\vert y\vert\!<\!\sqrt{\frac{\mathrm v_1}{\!1\!+\!\mathrm v_1\!}}$, decreasing in the region $y^2\in(\frac12,\frac{\mathrm v_1}{1+\mathrm v_1})$.

The equations of motion follow from the variations of the action with respect to $g^{\mu\nu}\!$ and $\varphi$:
\[
\phi_{,\mu}\phi^{,\nu}+\left(\mathrm V-\frac{\mathrm w^2}{2}\right)\!\delta_\mu^\nu=\frac12G_\mu^\nu,\quad x^{;\mu}_{;\mu}+m^2x\!\left(x^2\!-1\right)\!=0,
\]
where $\mathrm V_{(0,1)}=\frac{V_{(0,1)}}{m_P^2}$, $\mathrm w^2=\phi_{,\mu}\phi^{,\mu}$, and $G_\mu^\nu=R_\mu^\nu-\frac 12R\delta_\mu^\nu$. The equation $\varphi$ is contained in the equations of gravity due to the Bianchi identities ($G_{\mu;\nu}^\nu=0$). 

The Friedmann equations for a homogeneous field have the from:
%(14)
\begin{equation}
H^2=\frac{\dot\phi^2+2\mathrm V}{3}=\frac{2\mathrm V}{3-\gamma},\quad
\dot H=-\dot\phi^2,
\end{equation}
whence, we derive the following equations of motion for $\phi=\phi_1x$ and/or $\dot\phi=H\beta$:
%(15)
\begin{equation}
\ddot x+3H\dot x+m^2x\!\left(x^2\!-\!1\right)\!=0,\;\;\,
\frac{\beta\beta_{,\phi}}{3\!-\!\beta^2}+\beta=\frac{\mathrm n_0y\phi}{\mathrm v},\!
\label{eqmain}
\end{equation}
\[
H\!=\!H_0\sqrt{\!\frac{\mathrm v}{1\!-\!\frac{\gamma}{3}}}\!=\!\frac{2\mathrm m}{3}\sqrt{\frac{\bf v}{1\!-\!\frac{\gamma}{3}}},\;\;\,
\gamma\!=\!\frac{6\dot x^2}{2\dot x^2\!+\!m^2v}
%=\frac{\left(3-\gamma\right)\dot z^2}{m_1^2x^2(1+z^2)}
\in\left(0,3\right)\!,
\]
where $H_0\!=\!\sqrt{\frac{2\mathrm V_0}{3}}\!=\!\frac{m}{\sqrt{3\mathrm n_0}}$, $\mathrm m=\sqrt2m$, ${ \bf v}_{(1)}=\frac{3\mathrm  v_{(1)}}{8\mathrm n_0}$,
$H_1=\sqrt{\frac{2\mathrm V_1}{3}}=H_0\sqrt{\mathrm v_1}=\!\frac{2\mathrm m}{3}\sqrt{{\bf v}_1}$. 
The equation for $x$ can be rewritten in terms of the variable $\mathrm x=1-x$, which is convenient to use to describe the solutions of the residual vacuum:
%(16)
\begin{equation}
\ddot{\mathrm x}+3H\dot{\mathrm x}+m^2\mathrm x\left(1-\mathrm x\right)\left(2-\mathrm x\right)=0.
\label{eqx}
\end{equation}
Let us rewrite Eq.~(15) for the variables $X=xe^{\frac32N}$ and $\mathrm X=\mathrm xe^{\frac32N}$:
%(17)
\begin{equation}
\ddot X-\mathrm m^2{\bf v}_+ X=0,\quad
\ddot{\mathrm X}+\mathrm m^2{\bf v}_-\mathrm X=0,
\label{eqX}
\end{equation}
where 
${\bf v}_+=\frac y2+\bar{\bf v}$, 
${\bf v}_-=\frac{x(1+x)}{2}-\bar{\bf v}$, and
$\bar{\bf v}={\bf v}(\frac{3-2\gamma}{3-\gamma})$. 
The passage from $t$ to $mt$ results in the disappearance of the mass parameter ($m$ determining the time/energy) in the equations and evolution depends on the two constants $\mathrm n_0$ and $\mathrm v_1$.

The solutions of the equation for $\phi$ are trajectories on the $(\phi,\dot\phi)$ phase plane including three points (poles): central $\phi=\dot\phi=0$ and side $\phi\pm\phi_1=\dot\phi=0$, where $\gamma=\beta=\varepsilon_{,N}=0$, $H=H_{0,1}$, and $\varepsilon=\varepsilon_{0,1}$ respectively. In the central pole, two parameters $H_0$ and $\mathrm n_0$, while the third $\mathrm v_1$ is not included in the derivatives (\ref{g5}). In the side poles, all three parameters, $H_1$, $\mathrm v_1$, and $\mathrm n_1\!=\frac{2\mathrm n_0}{\mathrm v_1}=\frac{3}{4{\bf v}_1}$ are given. The constants $\varepsilon_{0,1}$ are related to the vacuum constants $\mathrm n_{0,1}$ through two independent binomials (see Eq.~(\ref{eqmain})):
%(18)
\begin{equation}
\varepsilon_0^3+3\varepsilon_0-3\mathrm n_0=0,\quad \varepsilon_1^2+3\varepsilon_1+3\mathrm n_1=0.
\end{equation}
The first of Eqs. (18) at $\varepsilon_0>0$ has the only solution
\[
\varepsilon_{0+}\equiv n_0=\frac{3\mathrm n_0}{\vert\varepsilon_{0-}\vert}=\frac{2\mathrm n_0}{1+\sqrt{1+\frac{4\mathrm n_0}{3}}},
\]
where $\varepsilon_{0\pm}=\frac32(-1\pm\sqrt{1+\frac{4\mathrm n_0}{3}})$. 
The second root $\varepsilon_{0-}$ is negative because the binomial is invariant under the transformation $\varepsilon_0\rightarrow-3-\varepsilon_0$. At ${\bf v}_1>1$ ($\mathrm n_1<\frac34$) both roots 
$\varepsilon_{1\pm}=\frac32(-1\pm\sqrt{1-{\bf v}_1^{-1}})$ of the second of Eqs.~(18) are negative, which will be used to obtain the VA solution\footnote{Two trajectories enter/exit in/from the central pole along the $\dot\phi=-(3+n_0)H_0\phi$ and $\dot\phi=n_0H_0\phi$ axes, respectively. The other trajectories, approaching the point $\dot x=x=0$, turn without entering the pole (which indicates instability). All trajectories, except for the central point, enter the side (stable) poles.}.

The solution of Eqs. (15) was sought in the form
%(19)
\begin{equation}
\phi=C\exp\left(\int ndN\right)\!,\quad 
\beta=n\phi,\quad\varepsilon=\left(nx\right)_{,x},
\end{equation}
where $C$ is a constant. The equation for $n=n(x)$ has the form
\[
n_{,N}=nn_{,x}x=\left(3-\gamma\right)\!\left(\mathrm n-n\right)-n^2\!=
\!\left(n_+-n\right)\!\left(n-n_-\right)\!,
\]
where $\mathrm n=\frac{\mathrm n_0y}{\mathrm v}$, and
$n_{\pm}=\frac{3-\gamma}{2}\left(-1\pm\sqrt{1+\frac{4\mathrm n}{3-\gamma}}\right)$.
The functions $n$ and $\mathrm n$ are related as 
%(20)
\begin{equation}
n=\frac{\mathrm n}{1+\frac{\epsilon}{3}},\qquad\epsilon=\frac{\varepsilon}{1-\frac{\gamma}{3}}.
\end{equation}
Here, $\epsilon$ satisfies the equation
\[
\epsilon_{,N}\!=
\!3{\bf n}+\!\left(\gamma\!-\!3\right)\!\left(\epsilon\!+\!\frac{\epsilon^2}{3}\right)\!=
\!\left(1\!-\!\frac{\gamma}{3}\right)\!\left(\epsilon_+\!-\!\epsilon\right)\!\left(\epsilon\!-\!\epsilon_-\right)\!,
\]
where ${\bf n}\!=\!(\mathrm nx)_{,x}\!=\!\frac{\mathrm n_0{\mathrm y}}{\mathrm v}$, 
$\mathrm y=\mathrm v(\frac{xy}{\mathrm v})_{,x}\!=1+\frac{x^2(\mathrm v-4\mathrm v_1)}{\mathrm v}$, $\epsilon_{\pm} = \frac{\varepsilon_\pm}{1-\frac{\gamma}{3}} = \frac 32\left(-1\pm\sqrt{1+\frac{4{\bf n}}{3-\gamma}}\right)$. 

The function $\mathrm y$ in the segment $x\in\!(0,1)$ varies from 1 to $-2$, passing through zero ($\mathrm y_*=0$) at the point $x_*^2=\frac{ly_*}{2}=\frac{l}{l+2}\in\!(\frac13,1)$:
%(21)
\begin{equation}
\mathrm y=\frac{\hat y\left(c_0c_1\!+\!\left(l-1\right)\tilde y-\tilde y^2\right)}{\left(l+2\right)\left(c_1+2\tilde y\!+\tilde y^2\right)}=c\hat y+\!O\!\left(\hat y^2\right)\!,
\end{equation}
where $\tilde y=\frac{y-y_*}{y_*}$ and $\hat y=2\tilde y$ are variables, $c=\frac{c_0}{l\!+\!2}\in\!(0.7,1)$, the constants are related by the parameter $l>1$,
\[
c_0=l+1+l^{-1}\!,\;\; c_1=\frac{2l}{l-1},\;\;\mathrm v_1=\frac{\mathrm v_*\left(l+1\right)}{2l}=\frac{4\left(l+1\right)}{l\ell},
\]
\[
v_1\!=\frac{y_*^2\!\left(l\!+\!1\right)\!}{l-1}\!,\;\,\ell\!=l^2\!+\!3l\!+\!4\!=\!\left(l_+\!+\!2\right)\!\left(l_-\!+\!2\right)\!,\;\,l_\pm\!=l\pm\sqrt{l}.
\]

The approximation of the general solution has the form
\[
n=\!\frac{n_0y}{u\mathrm v},\;\;\beta=\!\frac{\beta_1xy}{u\mathrm v},\;\;\epsilon=3\!\left(u\!-\!1\right)+n_0u=\!\frac{n_0\!\left(\mathrm y\!-\!\frac{u_{\!,x}xy}{u}\right)\!}{u\mathrm v\!\left(1\!-\frac{\gamma}{3}\right)},
\]
where $u=\!\frac{3+\epsilon}{3+n_0}$, $\beta_1=\sqrt{\gamma_1}=n_0\phi_1=\sqrt{\frac{2\bf n_0\!}{v_0}}$, and ${\bf n}_0\!=\frac{3n_0}{3+n_0}$.

The VA is a partial solution of the Friedmann equations, consisting of a sequence of stages: the motion occurs from the quantum boundary to the central pole, approaches it along the axis $\dot\phi=-(3+n_0)H_0\phi$ and exits along $\dot\phi=n_0H_0\phi$, where $n=\epsilon= \varepsilon= n_\pm= \epsilon_\pm= (n_0,-3-n_0)$ near $\phi=\dot\phi=0$. The exit of the VA from the center into the region $\phi>0$ corresponds to the observational model given by Eq. (9) with $\xi=\frac{6x^2}{v_0(3+n_0)}$, where the constants $H_0$ and $n_0$ are identical to those described in (9). Using the smallness $n_0$, we obtain the functions of $x\in(0,1)$ in the form of a series on 
$\zeta=x\bar\zeta=\frac{{\bf n}_0x^2}{\mathrm v}$\,\footnote{The constants $n_0$, $\mathrm n_0$, and ${\bf n}_0$ are known from observations, $\sqrt{\mathrm n_0{\bf n}_0}=n_0$, $\sqrt{\mathrm n_0/{\bf n}_0}=1+\frac{n_0}{3}$. The squares of the variables $\zeta$ and $\bar\zeta$ are smaller than each of them: 
$\zeta^2<\zeta\bar\zeta< \bar\zeta^2<\zeta<\bar\zeta<1$.} with $\gamma<1$:
\[
u=1+\mathrm u\zeta+\frac{{\bf u}\bar\zeta^2\!}{3}+O\!\left(\frac{n_0^3x^2}{\mathrm v^3}\right)\!,\quad\gamma=\frac{2\zeta y^2}{u^2\!\left(y^2+v_1\right)},
\]
%(22)
\begin{equation}
\epsilon=\frac{n_0}{\mathrm v}\!\left(\mathrm y\!+\!{\bf u}\zeta\!+\!O\!\left(\bar\zeta^2\right)\!\right)\!\!,\;\;\phi\simeq\bar{\phi_0}y^{\mathrm v_1\!/2}\!\left(\frac{k\,e^{\phi^2\!/2\!}}{k_0\sqrt{\mathrm v}}\right)^{\!n_0}\!\!\!,
\label{eq35}
\end{equation}
\[
N\!+\!N_0\!\simeq\!\frac{1}{n_0\!}\!\left(\ln\!\vert x\vert\!-\!\frac{\mathrm v_1}{2}\!\ln\!\vert y\vert\!+\!\frac{\!2y\!-\!1\!+\!\ln2\!}{2v_0}\right)\!\!=\!\ln\!\left(\!\frac{k}{k_0\sqrt{\mathrm v}}\!\right)\!\!,\;
\]
where the functions $\mathrm u$ and $\bf u$ are bounded\,\footnote{These functions are given by the expressions:
\[
u =1+\!\frac{\epsilon-n_0}{3+n_0},\quad
\mathrm u=\frac{\mathrm y-\mathrm v}{3x^2}=1-\frac{2\mathrm v_1\left(\mathrm v+2\right)}{3\mathrm v}-\frac{\mathrm n_0\phi^2}{6},
\]
\[
{\bf u}=\mathrm y\bar{\mathrm u}+\frac{2{\mathrm y}_{,y}y}{3},\quad\bar{\mathrm u}=\frac13\left(-5+\frac{2\mathrm v_1\left(\mathrm v+3\right)}{\mathrm v}+\frac{\mathrm n_0\phi^2}{2}\right).
\]
}, 
as $\vert\mathrm u\vert<1$ and $\vert{\bf u}\vert<1$, the condition $k=\frac{k_0\sqrt{1+3\mathrm v_1}}{2}$ ($N=-N_0$) is fulfiled at the point $y\!=\!x^2\!=\!\frac12$ ($\phi=\phi_0$), the constants are related as $\bar{\phi_0}=\phi_1(\sqrt{2e})^{\mathrm v_1-1}$, $N_0\!=\!\ln(\frac{H_0}{k_0})$, $\phi_0=\!\frac{\phi_1}{\sqrt2}\!=\!\frac{1}{\sqrt{v_0n_0}}\!\simeq\!7\sqrt{1\!-\!\mathrm v_1}$, and $\phi_1\!\simeq\!10\sqrt{1\!-\!\mathrm v_1}$. 
The series of the VA are valid at $x<\frac89$ and $y>\frac29$. At smaller $y$, when the VA solution was near the pole $x\!-\!1\!=\!\dot x\!=\!0$, it is necessary to use Eq.~(\ref{eqx}).

The VA includes three constants $H_0$, $n_0$, and $\mathrm v_1$\,\footnote{The potential of all the fields responsible for the polarization of the vacuum is obtained as a series of fields near zero. At the first stage of the CVR, the first three terms of the series of the dominant field were taken, without increasing the number of potential constants (the higher terms were assumed to be negligible).
The three coefficients of the three terms of the series (\ref{3const}) give the three constants of the VA. The method of reducing the cosmological constant by introducing a ``specific'' interaction potential is not the topic of our work. The potential of the model \cite{Abbott85} contains a larger number of constants and higher derivatives of the field (in the form of a cosine field), while the passage from large to small fields does not correspond to our approach (see footnote 5). We treat the vacuum (the ground or zero state of all fields) as its polarization in the external gravitational field with the subsequent relaxation of the vacuum in time as a generator of the evolving Universe.}. 
The first two constants are determined from observations (see Eq.~(\ref{eq10})), and the third constant is free: it does not affect the beginning of the VA and manifests itself later at $k>10$, when the solution changes. The functions $n$ and $\varepsilon$ are first (when $\phi<3$), close to each other and then (when approaching $\phi_1$) diverge depending on the ratio $\mathrm v_1$ and $\mathrm v_{cr}=\frac{8\mathrm n_0}{3}\simeq0.05$, ${\bf v}_1=\frac{\mathrm v_1}{\mathrm v_{cr}}\simeq20\mathrm v_1$ (see Fig.~2).

%%%%%%%%%%%%%%%%%%%%%%%%%%%%%%%%
\begin{figure}
\centering
\includegraphics[width=0.5\textwidth]{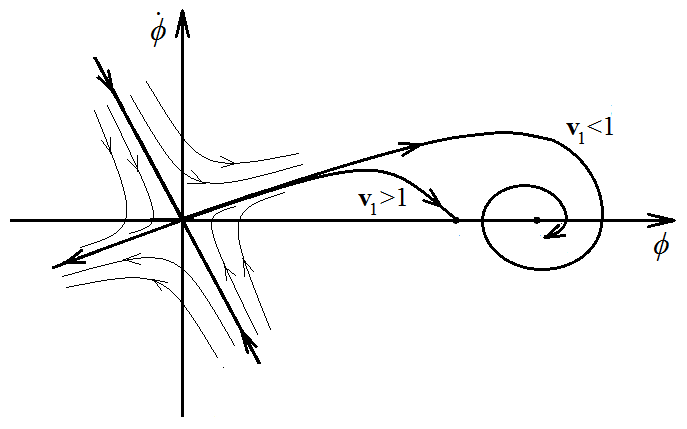}
\caption{Trajectories of the solution of Eq.~(\ref{eqmain}) on the 
($\phi$, $\dot\phi$) phase plane. Thick and thin lines correspond 
to the vacuum attractor and the general solution, respectively, and
arrows indicate the direction of time.}
\label{fig2}
\end{figure}
%%%%%%%%%%%%%%%%%%%%%%%%%%%%%%%
%%%%%%%%%%%%%%%%%%%%%%%%%%%
\subsection{Case ${\bf v}_1<1$ ($\mathrm v_1<0.05$).}
%%%%%%%%%%%%%%%%%%%%%%%%%%%
The function $\gamma$ increases monotonically with the
time, approaching $\gamma\sim1$ at $t\simeq\mathrm m^{-1}$ ($x\simeq0.87$, $y\simeq\sqrt{\mathrm v_{cr}}\simeq0.2$, $H\simeq\frac{2\mathrm m}{3}$). At $t>\mathrm m^{-1}$ ($y<0,2$), the VA trajectory passes clockwise around the pole, crossing the $x=1$ and $\dot x=0$ axes at the times $t_1$ and $t_b>t_1$, respectively, passing from the upper to lower part of the phase plane. Further, the field continues to oscillate around the point $y=\dot y=0$, approaching it: $y\simeq0.2\,\frac{\cos(\mathrm mt)}{\mathrm mt}$ ($H\simeq\frac{2}{3t}$) at $t\in(\mathrm m^{-1},t_{\mathrm v})$, and $y\simeq e^{-t/t_{\mathrm v}}\cos(\mathrm m_{\mathrm v}t)$ ($H\simeq\frac{2}{3t_{\mathrm v}}$) at $t>t_{\mathrm v}$, where $t_{\mathrm v}=\frac{1}{\mathrm m\sqrt{{\bf v}_1}}$, $\mathrm m_{\mathrm v}=\mathrm m\sqrt{1-{\bf v}_1}$, and the phases $\sim1$ in cosines are omitted. Thus, the first stage of the cascade ends, and the problem is reduced to the previous one: the second stage of the CVR begins with a new vacuum $V=V_1$ and a new dominant field.

%%%%%%%%%%%%%%%%%%%%%%%%%%%%%%%%%%%%%%%%%%%%%%%
\subsection{Case ${\bf v}_1\in (1,20)$ ($\mathrm v_1=\frac{4(l+1)}{l\ell}\in(\frac{1}{20},1)$, 
$l\in (1,8)$, $v_1=\frac{4(l+1)}{(l+2)^2(l-1)}>\frac{1}{20}$, $\sqrt{\mathrm v_1}\in(\frac29,1)$).} 
%%%%%%%%%%%%%%%%%%%%%%%%%%%%%%%%%%%%%%%%%%%%%%%
The function $\gamma$ increases with the field, reaches a maximum $\gamma_r\!=\!\beta_r^2\!\simeq\!\frac{n_0(l\!-\!1)(l\!+\!2)}{9}\!\simeq\!\frac{0,007}{v_1}$ at $x_r\!\simeq\!\sqrt{\frac{l}{l\!+\!2}}\!\in\!(0.6,0.9)$ and returns back to zero at $x\rightarrow1$, remaining small in the region $x\in(0,1)$. At $x\in(0,\frac89)$ and $y\in(1,\frac29)$ we have $\beta\simeq\frac{2xy}{\phi_1v}\le\beta_r\simeq\frac{0.08}{\sqrt{v_1}}\in(0, 0.4)$ and $\dot x\simeq\frac{\mathrm mxy}{\sqrt{3v}}$. In the region $\mathrm x\in(\frac19,\frac12\sqrt{\mathrm v_1})$ we have $\beta\simeq\frac{4\mathrm x}{\phi_1v_1}$ and $\dot x\simeq\frac{2\mathrm m\mathrm x}{\sqrt{3v_1}}$. When $\vert\mathrm x\vert<\frac12\sqrt{\mathrm v_1}$ the solution of (\ref{eqx}) has the form
\begin{equation}
\mathrm x\!=\!\mathrm x_{\!\times\!}\!\left(\!\mathrm c_+\!e^{\!-\!\bar t_+}\!\!+\!\mathrm c_-\!e^{\!-\!\bar t_-\!}\!\right)\!=\!\mathrm x_{\!\times\!}e^{\!-\!\bar t\!}\left(\!\text{ch}\left(\omega\bar t\right)\!-\!\frac{\mathrm c\cdot\text{sh}\!\left(\omega\bar t\right)\!}{\omega}\right)\!\!,
\end{equation}
\[
\frac{\dot x}{\mathrm m\sqrt{{\bf v}_1}\mathrm x}=1+\omega+\!\frac{2\omega}{{\bf c}e^{-2\omega\bar t}\!-\!1}=1+\mathrm c+\!\frac{\left(\mathrm c^2\!-\!\omega^2\right)\!\text{th}\!\left(\omega\bar t\right)\!}{\omega-\mathrm c\cdot\text{th}\!\left(\omega\bar t\right)\!},
\]
where $\bar t=\mathrm m\sqrt{{\bf v}_1}(t-t_{\times})\geq0$ and $\bar t_{\pm}=(1\pm\omega)\bar t$ are variables, $\mathrm x_\times$, $\omega=\sqrt{1-{\bf v}_1^{-1}}$, $\mathrm c_\pm=\frac{\omega\pm\mathrm c}{2\omega}$, ${\bf c}=\frac{\mathrm c+\omega}{\mathrm c-\omega}$, and $\mathrm c$  are constants. The quantities $\mathrm x_{\times}\!\in\!(\frac19,\frac12\sqrt{\mathrm v_1})$ and $\mathrm c\simeq\frac{2}{\sqrt{3v_1{\bf v}_1}}-1\in(4.2,-1)$ are derived from the matching conditions with the VA at  $\bar t=0$ ($\mathrm x=\mathrm x_{\times}$, $\dot{\mathrm x}\simeq\frac{2\mathrm m}{\sqrt{3v_1}}\mathrm x_{\times}$).

We consider two cases.

At $t>t_\times$ and $\mathrm c>\omega$ ($v_1<\frac17$, $\mathrm v_1<\frac18$, ${\bf v}_1\in (1,2.7)$, $\omega\in(0,0.8)$, $l\in(4.5,8)$)\footnote{The condition $c^2=\omega^2$ leads to the equation 
$\frac{4}{3v_1}+1=\phi_1\simeq\frac{10.7}{\sqrt{v_0}}$, from which we obtain $v_1\simeq\frac17$ and $\omega\simeq\frac45$ (we omitted the second root $v_1\simeq100$).}, the VA trajectory passes clockwise around the pole $\mathrm x=\dot{\mathrm x}=0$, crossing the $x=1$ axis in the upper half of the phase plane at $\bar t_1=\frac{\ln{\bf c}}{2\omega}$ and crossing $\dot x=0$ axis towards the lower half of the phase plane at $\bar t_b$, and then at $\bar t>\bar t_2$ approaches the pole radially from the right ($\phi>\phi_1$) along the $\dot\phi=\frac{\mathrm m(\phi_1-\phi)}{\sqrt{{\bf v}_1}+\sqrt{{\bf v}_1-1}}$ axis, where $\bar t_{1,2}=\bar t_b\pm\Delta_b$ and $\Delta_b=\frac{\ln (\frac{1+\omega}{1-\omega})}{2\omega}=\frac{\ln(\sqrt{{\bf v}_1}+\sqrt{{\bf v}_1-1})}{\omega}$. At this point, the function $n$ approaches zero, and the function $\varepsilon$ is $\varepsilon\simeq(\ln\vert\dot x\vert)_{,N}$. The latter function is constant at $\bar t<\bar t_1$ ($\varepsilon\simeq-n_1$) and $\bar t>\bar t_2$ ($\varepsilon\simeq-n_2$, where $n_2=\frac{2n_1}{1+\omega}$)\,\footnote{ From (23), we obtained
\[
\mathrm x\!\propto\!\left({\bf c} e^{-\tilde t}\!-\!1\right)\!e^{-\bar t_-}\!,\;\;\dot x\!\propto\!\left(e^{\tilde t_b-\tilde t}\!-\!1\right)\!e^{-\bar t_-}\!,\;\;\ddot x\!\propto\!\left(\!1-\!e^{\tilde t_2-\tilde t}\right)\!e^{-\bar t_-}\!,
\]
where $\tilde t=2\omega\bar t$, $\tilde t_b=\ln({\bf c})+
2\ln(\sqrt{{\bf v}_1}+\!\sqrt{{\bf v}_1-1})$, $\tilde t_{1,2}=\tilde t_b\pm2\ln(\sqrt{{\bf v}_1}+\sqrt{{\bf v}_1-1})$.
}, but diverges in the time interval $\bar t\in(\bar t_1,\bar t_2)$, being an index (derivative of the logarithm) of the function $H_1^2/\dot\phi$ in the form of a high peak ($\delta$-function): $\varepsilon\simeq(\ln\vert\dot x\vert)_{,N}$, which gives a local feature (bump) in the spectrum $q_k$ at $k_b=H_1a(t_b)$. 

At $\mathrm c<\omega$ ($v_1>\frac17$, ${\bf v}_1>e$ and $l\in(1,4.5)$), the VA trajectory approaches the $x=1$ pole radially from the left ($\phi<\phi_1$) along $\dot\phi\simeq \frac{\mathrm m(\phi_1-\phi)}{\sqrt{{\bf v}_1}+\sqrt{{\bf v}_1-1}}$ for a formally infinite time.

%%%%%%%%%%%%%%%%%%%%%%%%%
\section{NON-POWER-LAW POWER SPECTRA}
%%%%%%%%%%%%%%%%%%%%%%%%%
At $k\ll k_0$, the power spectrum has a power-law form decreasing with an increase in the wavenumber $k$ in accordance with observations (see Eq.~(\ref{eq35})):
%(23)
\begin{equation}
q_k\!=\!\bar q_0\!\left(\frac{k_0}{k}\right)^{\!n_0}\!\!,\;\,r_k\!=\!\bar r_0\!\left(\frac{k}{k_0}\right)^{\!2n_0}\!\!,\;\,\phi\!=\!\bar{\phi_0}\!\left(\frac{k}{k_0}\right)^{\!n_0}\!\!,
\label{asi}
\end{equation}
where $\bar q_0\!=\!\frac{\mathrm H_0}{2\pi n_0\bar\phi_0}\!\simeq\!10^{-\!5}(\frac{k_c}{k_0})^{n_0}\!$, $\bar r_0\!=\!4n_0^2\bar\phi_0^2$, and $\bar{\phi_0}\!=\!\phi_c(\frac{k_0}{k_c})^{n_0}\!$.  
The continuation of the spectrum to scales of hundreds of kiloparsecs and less depends on
the third constant ${\bf v}_1$, resulting in additional power in the form of a ``bump'' and/or a blue spectrum of density perturbations.

At ${\bf v}_1<1$, the ratio of the spectra $r_k$ continues to increase with $k$ and becomes large ($\sim1$) at $k= k_r\sim k_1$. Soon, at $k=k_p>k_1$, an intense peak appears in the spectra $q_k$, leading to the production of primordial black holes. At higher wavenumbers $k>k_p$ the spectra decrease due to oscillations of the field $\phi$ near the side pole. At ${\bf v}_1\ll1$ oscillations lead to instability: the creation of field particles with the mass $\mathrm m$ and their further decay with the transition to the hydrodynamics of radiation-dominated plasma. This certainly refers to the decay of the oscillation part of the field $\varphi$, but all fields included in the polarization of the vacuum remain stable (in particular, the state $V_1$ remaines the same). 

At ${\bf v}_1>1$ and $x\in(0,0.9)$, the spectra are two-power:
%(21)
\begin{equation}
q_k\!=\frac{q_0\mathrm v^{\frac32}}{xy},\;
r_k\!=r_0\!\left(\frac{xy}{\mathrm v}\right)^{\!2\!}\!,\;\hat 
x^{\frac{1}{n_0}}\hat y^{-\frac{1}{n_1}}\!=\frac{k\,e^{\frac{\phi^2\!-\!\phi_{01}^2}{2}\!} }{k_0\sqrt{\mathrm v}},
\label{eq39}
\end{equation}
\[
n_k\!\simeq\!-\frac{n_0}{\mathrm v}\!\left(1\!+\!3x^2\!\left(1\!-\!2\frac{\mathrm v_1}{\mathrm v}\right)\!\right)\!\in\!\left(-n_0,n_1\right)\!,\;\;n_1\!=\!\frac{2n_0}{\mathrm v_1},
\]
where $q_0=\bar{q_0}(\sqrt{2e})^{\mathrm v_1-1}$, 
$r_0\!=\bar{r_0}(2e)^{1-\mathrm v_1}$,
$\hat x=\frac{\phi}{\phi_{01}}\! =\!\sqrt2x$, 
$\hat y=2y$, and $n_1\in(0.04,0.7)$. 
At $k<k_0$ ($x<0.6$), Eqs.~(\ref{eq39}) correspond to asymptotic Eqs.~(\ref{asi}). At $k>k_0>k_1$ ($y\in(0.2,\sqrt{\frac{\mathrm v_1}{1+\mathrm v_1}})$), the spectra increase with
the wavenumber $k$:
\begin{equation}
q_k\!=\!q_1\!\left(\frac{k}{k_1}\right)^{\!n_1\!}\!\!,
\;\,r_k\!=\!r_1\!\left(\frac{k_1}{k}\right)^{\!2n_1\!}\!\!,
\;\,
x^2\!=\!1\!-\!\left(\frac{k_1}{k}\right)^{\!n_1\!}\!\!,
\label{eq40}
\end{equation}
where $q_1\!=q_0\mathrm v_1^{3/2}$, $r_1\!=\frac{r_0}{\mathrm v_1^2}$, and $k_1=k_0\sqrt{\mathrm v_1}(\sqrt{\frac2e})^{\phi_{01}^2}$. Formulas (\ref{eq39}) are arithmetically simple and do not require any approximation. 

A further increase in $q_k$ depends on the residual vacuum. At $\vert x\vert<1/2$, the field equation has the form 
\[
\ddot\delta +\mathrm{m}\sqrt{4V_1+3\delta^2}\dot\delta+\mathrm{m}^2\delta=0,
\]
where $\delta=\phi-\phi_1=-\phi_1\mathrm{x}$. 
At ${\bf v}_1\in(1,e)$, a spike (bump) appears in the spectrum at the wavenumber
$k_b$, which has the form of a single peak with the amplitude $q_b\,{}^<_\sim\, 1$\,\footnote{The peak is approximated by the Gaussian, while the spectrum returns to power-law evolution at $k>k_b$. Thus, the power release at $k\simeq k_b$ is the bump on the increasing part of the three-power spectrum $q_k$ with the indices $n_k=(-n_0,n_1,n_2)$.}. When \textbf{v}${}_1\in(e,e^3)$, there is no bump in $q_k$, and the spectrum has a three-power form with the indices $n_k=(- n_0, n_1, n_2)$, where $n_2\simeq n_1$. At the end of the first stage of the CVR, the field is frozen at the pole in the state $V_1$ for a formally unlimited time, $\phi\rightarrow\phi_1$. From this state $V_1$, the second stage of the CVR will begin with a new field (etc., on the cascade of $V$ potential steps). 

These changes in the spectrum at $k > 10$Mpc$^{-1}$ lead both to the early formation of stars (as indicated by James Webb Space Telescope data) and to the production of primordial black holes and the early occurrence of supermassive black holes (as indicated by LIGO data and \cite{SMBH}, respectively). The spectrum can be continued to a small scale if conclusions based on the James Webb Space Telescope data and studies of the mass function of dark matter halos and the evolution of subhalos in large galaxies will be confirmed (see \cite{tkachev24a, eroshenko24, tkachev24b} and references therein). Subsequent more accurate observation data will capture the characteristics of the power spectrum at small scales.

%%%%%%%%%%%%%%%%%%%%%%%%%%%%%%%%%%%%%%%%
\section{ENTERING VACUUM ATTRACTOR IN $V_0$ AND LEAVING IT IN $V_1$}
%%%%%%%%%%%%%%%%%%%%%%%%%%%%%%%%%%%%%%%%
Near the central pole $\vert\phi\vert<\phi_0$ and $\vert\dot\phi\vert<H_0$, the solution of Eq.~(\ref{eqmain}) is the sum of rising and falling time exponentials:
%()
\begin{equation}
\phi= C_0 a^{n_0}+\bar C_0 a^{-\bar n_0},\quad a=\frac{k}{H_0}\propto e^{H_0t},
\end{equation}
where $\bar n_0\! =\! 3 +n_0$, the constants $C_0$ and $\bar C_0$ are proportional to $C$, and their ratio is included in the scale of the exponential equality $k_i\!=\!H_0a_i\!$ (beginning/boundary/entering into the VA):
\[
\phi = \phi_i\kappa^{n_0}\left(1\pm\nu^2\kappa^{-n_*}\right),
\]
%(28)
\begin{equation}
 n =n_0\frac{1\mp\nu\kappa^{-n_*}}{1\pm\nu^2\kappa^{-n_*}},\quad
 \varepsilon=n_0\frac{1\pm\kappa^{-n_*}}{1\mp\nu\kappa^{-n_*}},
\end{equation}
where $\kappa=\frac{k}{k_i}=e^{H_0(t-t_i)}$, $n_*=n_0+\bar n_0=\sqrt{3(3+4\mathrm n_0)}$, and $\nu=\frac{n_0}{\bar n_0}\simeq0.005$. The upper sign refers to the right ($\phi_i>0$) or left ($\phi_i<0$) quadrants of the phase plane bounded by the VA, and the lower sign refers to the upper or lower quadrants.

Coming to the vicinity of the central pole, the field trajectories are under the gravitational influence of the giant swirl of the pole, which draws the trajectories to the partial self-similar VA solution at $\kappa\sim1$. Approaching the pole from the side of $\vert\beta\vert\sim1$, the trajectories move along the VA, approaching the VA at $\kappa>1$, and turn near the central pole, moving away from it along with the VA to the side of $\vert\phi\vert\sim\phi_1/2$:
%(29)
\begin{equation}
n=n_0\left(1\mp\frac{\nu\!\left(1+\nu\right)\!}{\kappa^{n_*}}+O\!\left(\frac{\nu^3}{\kappa^{2n_*}}\right)\!\right)\!,
\end{equation}
\[
\varepsilon=n_0\!\left(1\pm\frac{1+\nu}{\kappa^{n_*}}+O\!\left(\frac{\nu}{\kappa^{2n_*}}\right)\!\right)\!.
\]
Only the growing exponential remains in the evolutionary trajectory of the solution, since the falling exponential quickly becomes insignificant. This means the entry of the solution into the VA near $V_0$ in the initial interval with $\phi\in(\phi_i,\phi_0)$, $\kappa\in(1,e^{N_{0i}})$, and constant indices $n\simeq\varepsilon\simeq n_0$ (cf.~Eq.~(22)):
\[
H=H_0\left(1-\frac{n_0\phi^2}{2}\right),\quad\phi=\phi_i\kappa^{n_0}\simeq\phi_0e^{n_0\left(N+N_0\right)},
\]
where $N_{0i}\simeq57 \ln(\frac{\phi_0}{\phi_i})$. The index of the power spectrum is weakly dependent on $\phi$ ($q_k\propto H^2/\phi$, $n_k\simeq-n_0$) and agrees well with the observational model given by Eq.~(9) with $\phi_c\in(\phi_i,\phi_0)$, $k_0\simeq0.05 \exp{(N_{0c})}$ Mpc$^{-1}$, and $N_{0c}\simeq57\ln(\frac{5\sqrt{1-\mathrm v_1}}{\phi_c})$. The continuation of the VA towards $V_1$ depends on the only free parameter $\mathrm v_1$, which is not yet limited by modern data.

Near the side pole, at $\vert\mathrm x\vert<0.5$, $\vert\dot\phi\vert<H_1$, and ${\bf v}_1>1$, the solution of Eq.~(\ref{eqx}) is the sum of two falling exponentials:
%(30)
\begin{equation}
\phi=C_1 a^{-\bar n_1}+C_2 a^{-\bar n_2},\quad a=\frac{k}{H_1}\propto e^{H_1t},
\end{equation}
where $\bar n_{1,2}=-\varepsilon_{1\pm}=\frac32(1\mp\omega)$. As seen, the first stage of the VA brings the field to one of the side poles with $\phi=\pm\phi_1$ and $\dot\phi=0$, where the field is in a thermal bath formed by quantum fluctuations of all fields, i.e., in a thermostat with a temperature of $T=\frac{H_1}{2\pi}$, for a formally unlimited time \cite{GibbonsHawking, Starobinsky, Volovik}. This situation differs from the de Sitter spacetime, since the landscape $V$ is not a constant (the state $V_1$ is stable for $\phi$, but not for all fields). Consequently, there is always a field (not $\phi$), that will continue the CVR from $V_1$ to a lower state $V_2<V_1$. The CVR itself is a VA: it can be reached once, as in the state $V_0$ for our trajectory of the Universe, and the CVR, as well as the VA, stops when the landscape ends.   

%%%%%%%%%%%%%
\section{CONCLUSIONS}
%%%%%%%%%%%%%
To summarize, we have build a solution based on the recent observational data. This solution be called ``elegant'', because it uncouples three known independent parameters: the spectral index, the ratio of power spectra, and the number $N$ of e-folds. The discussed epoch of the increase in the field $\phi$ from zero was characterized by a small ratio of power spectra $r_k<0.01$ and a constant index ($n_k\simeq-0.02$) of the spectrum of density perturbations. As long as $r_c$ is unknown, the observation scales can refer to any time with $\varphi_c<0.8\,M_P$. The observational model of vacuum relaxation is based on current data and provides a power-law red spectrum at $k<10$, without requiring information of the potential. The theoretical model of the vacuum attractor has two known constants, and the third constant ${\bf v}_1$ is needed to continue the vacuum attractor to $k>10$. The comparison of the vacuum attractor with the observational model of vacuum relaxation leads to the concept of the cascade vacuum relaxation as the generator of the evolving Universe, which solves the problem of the observed cosmology.

%%%%%%%%%%%%%%%%%%%%%%%%%%%%%%%%%%%%%%%%%%
\section*{APPENDIX. QUANTUM-GRAVITATIONAL PRODUCTION OF $S$ AND $T$ 
PERTURBATION MODES OF THE METRIC IN THE FRIEDMANN MODEL}
%%%%%%%%%%%%%%%%%%%%%%%%%%%%%%%%%%%%%%%%%% 
The expansion of the action $S=\int(-\frac{m_P^2}{4}R+\mathcal L)\sqrt{-g}d^4x$ in linear perturbations gives $S=S^{(0)}+\delta^{\left(1\right)}S+\delta^{\left(2\right)}S$, where $S^{(0)}=\int(-\frac32m_P^2H^2+\mathcal L^{(0)})a^3dtd\bf x$, $\delta^{\left(1\right)}S$ vanishes on the Friedmann equations, $\delta^{\left(2\right)}S=\int(L+\tilde L)a^3dtd\bf x$, and $L=\frac12\alpha^2(c_S^{-2}\dot q^2-q_{,i}q^{,i})$ is the Lagrangian density of the $S$ mode with the functions $\alpha^2=m_P^2\gamma$ and $c_S^{-2}=\frac{w\mathcal L_{,w,w}}{\mathcal L_{,w}}$ (see \cite{Lukash80a, Lukash80b}, \cite{book2010}). Under the condition given by Eq.~(2), $c_S=1$.

The quantum fields of the 4-coordinates $x^\mu=(t,x^i)$, $q$ and $\tilde q_{ij}$, and 4-momenta $k_\mu=(\frac ka,k_i)$,  $q_{\bf k}$, and $q_{{\bf k}\xi}$, in the Euclidean three-dimensional spaces ${\bf x}=(x^i)$ and ${\bf k}=(k_i)$, of the Friedmann model, respectively, are related to each other by the Fourier transform:
\[
q=\int\limits_{-\infty}^{\infty} q_{\bf k}e^{i{\bf k}{\bf x}}\frac{d\bf k}{\left(2\pi\right)^{3/2}},\quad q_{\bf k}=\nu_ka_{\bf k}+\nu^*_ka^{\dagger}_{-\bf k}, \eqno{(A.1)}
\]
\[
\tilde q_{ij}=\sum\limits_\sigma\int\limits_{-\infty}^{\infty} p_{ij\sigma} q_{{\bf k}\sigma}e^{i{\bf k}{\bf x}}\frac{d\bf k}{\left(2\pi\right)^{3/2}},\;\;
q_{{\bf k}\sigma}\!=\tilde \nu_ka_{{\bf k}\sigma}+\tilde \nu^*_k a^{\dagger}_{-{\bf k}\sigma},
\] 
where $p_{ij\sigma}=p_{ij\sigma}(\bf k)$ are the normalized constant polarization tensors of 
${\bf k}$-waves ($p^i_{i\sigma}=p_{i\sigma}^jk_j=0$, $p^*_{ij\sigma}p^{ij}_{\sigma^\prime}=\delta_{\sigma\sigma^\prime}=\text{diag}(1,1)$, $\sigma=\oplus,\otimes$).
The commutation relations between canonical conjugate scalars $q$ and $p=\frac{\partial L}{\partial\dot q}=\alpha^2\dot q$ in the position space are transferred to the commutators of the constant annihilation, $a_{\bf k}$, and creation, $a^\dagger_{\bf k}$, operators of particles (phonons) in the momentum space:
\[
\left[q\!\left(t,\!\bf x\right)p\!\left(t,\!{\bf x}^\prime\right)\right]\!=\!qp-pq\!=\!i\frac{\delta\!\left({\bf x}\!-\!{\bf x}^\prime\right)}{a^3},\;\left[a_{\bf k}a^\dagger_{{\bf k}^\prime}\right]\!=\!\delta\!\left({\bf k}\!-\!{\bf k}^\prime\right)\!,
\]
where $\delta=\delta(\bf x)$ is the Dirac delta function. The evolution of $q_{\bf k}$-oscillators is described by a classical normalized function $\nu_k=\bar\nu_k/(\alpha a)$ of the time $\eta=\int\frac{dt}{a}=\int\frac{dN}{aH}$ and the wavenumber $k=\vert\bf k\vert$:
\[
\bar\nu_k^{\prime\prime}+\!\left(k^2\!-U\right)\!\bar\nu_k\!=0,\quad U=\frac{\left(\alpha a\right)^{\prime\prime}}{\alpha a},\quad\bar\nu_k\bar\nu_k^{*\prime}- \bar\nu_k^* \bar\nu_k^\prime=i.
\]
Similar relations are valid for the annihilation, $a_{{\bf k}\sigma}$, and creation, $a^\dagger_{{\bf k}\sigma}$, operators of gravitons, and the evolution function $\tilde\nu_k=\sqrt2\bar{\tilde\nu}_k/(m_Pa)$ of $q_{{\bf k}\sigma}$-oscillators: 
\[
[a_{{\bf k}\sigma}a^\dagger_{{\bf k}^\prime\sigma^\prime}]=\delta\!\left({\bf k}\!-\!{\bf k}^\prime\right)\delta_{\sigma\sigma^\prime},\quad [a_{{\bf k}\sigma}a_{{\bf k}^\prime\sigma^\prime}]=0,
\]
\[
\bar{\tilde\nu}_k^{\prime\prime}+\!\left(k^2\!-\tilde U\right)\!\bar{\tilde\nu}_k\!=0,\quad\tilde U=\frac{a^{\prime\prime}}{a},\quad\bar{\tilde\nu}_k\bar{\tilde\nu}_k^{*\prime}- \bar{\tilde\nu}_k^* \bar{\tilde\nu}_k^\prime=i.
\]
The cosmological $S$ and $T$ modes of elementary oscillations of the Friedmann model are determined by the vacuum state $\vert0\rangle$ of all linear fields:
\[ 
a_{{\bf k}\left(\sigma\right)}\vert 0\rangle=\langle 0\vert a^\dagger_{{\bf k}\left(\sigma\right)}=0, \quad
\langle a_{{\bf k}}a^\dagger_{{\bf k}^\prime}\rangle=\delta\!\left({\bf k}\!-\!{\bf k}^\prime\right)\!,
\]
\[
\langle a_{{\bf k}\sigma}a^\dagger_{{\bf k}^\prime\sigma^\prime}\rangle=\delta({\bf k}\!-\!{\bf k}^\prime)\delta_{\sigma\sigma^\prime}.
\] 
The bilinear forms in Eqs.~(6) are given by the formulas (see Eq.~(A.1)): 
\[
q_k=\frac{k^{3/2}\vert\nu_k\vert}{\sqrt 2\pi},\quad\tilde q_k=\frac{k^{3/2}\vert\tilde\nu_k\vert}{\pi}, \quad r_k\!=\frac{\tilde q_k^2}{q_k^2}=\frac{2\vert\tilde\nu_k\vert^2}{\vert\nu_k\vert^2}.
\]

The parametric potentials of elementary oscillators have the form
\[
U\!=a^2\!H^2\!\left(\left(2\!-\!\gamma\!+\!\varepsilon\right)\!\left(1\!+\!\varepsilon\right)\!+\!\varepsilon_{,N}\right)\!,\;\;\,\tilde U\!=a^2\!H^2\!\left(2\!-\!\gamma\right)\!.
\]
The potentials coincide under the conditions
\[
\gamma\ll1,\; \vert\varepsilon\vert\ll1,\;\vert\varepsilon_{,N}\vert\ll1\!:\quad U=\tilde U=\frac{2}{\eta^2},\;\eta=\!-\frac{1}{aH},
\]
which solves the problem of cosmological perturbations in the Friedmann model:
\[
\bar\nu_k=\bar{\tilde\nu}_k=\frac{e^{-ik\eta}}{\sqrt{2k}}\left(1+\frac{1}{ik\eta}\right),\quad r_k=4\gamma,
\label{prilq}\eqno{(A.2)}
\]
\[
q_k=\frac{\sqrt{H^2+\frac{k^2}{a^2}}}{2\pi\vert\alpha\vert}\rightarrow\frac{\mathrm H}{2\pi\vert\beta\vert},\quad \tilde q_k=\frac{\sqrt{H^2+\frac{k^2}{a^2}}}{\pi m_P}\rightarrow\frac{\mathrm H}{\pi}.
\]
The passage to the limit means the time-frozen power spectra at wavelengths $k\ll aH$, where the functions $H=m_P\mathrm H$ and $\alpha=m_P\beta$ are calculated at the time $\eta=-1/k$.

\section*{ACKNOWLEDGMENTS}

We are grateful to the reviewers, whose comments
allowed us to clarify the text.

\section*{FUNDING}
This work was supported by the Russian Science Foundation,
project no. 23-22-00259.

\section*{CONFLICT OF INTEREST}
The authors of this work declare that they have no conflicts of interest.

\section*{OPEN ACCESS}
This article is licensed under a Creative Commons Attribution 4.0 International License, which permits use, sharing, adaptation, distribution and reproduction in any medium or format, as long as you give appropriate credit to the original author(s) and the source, provide a link to the Creative Commons
license, and indicate if changes were made. The images or other third party material in this article are included in the article's Creative Commons license, unless indicated otherwise in a credit line to the material. If material is not included in the article's Creative Commons license and your intended
use is not permitted by statutory regulation or exceeds the permitted use, you will need to obtain permission directly from the copyright holder.

%%%%%%%%%%%%%

%%%%%%%%%%
\end{document}